# Artificial Intelligence in 3GPP 5G-Advanced: A Survey

Xingqin Lin

NVIDIA

Email: xingqinl@nvidia.com

*Abstract*—Industries worldwide are being transformed by artificial intelligence (AI), and the telecom industry is no different. Standardization is critical for industry alignment to achieve widespread adoption of AI in telecom. The 3rd generation partnership project (3GPP) Release 18 is the first release of 5G-Advanced, which includes a diverse set of study and work items dedicated to AI. This article provides a holistic overview of the state of the art in the 3GPP work on AI in 5G-Advanced, by presenting the various 3GPP Release-18 activities on AI as an organic whole, explaining in detail the design aspects, and sharing various design rationales influencing standardization.

## I. INTRODUCTION

The 3rd generation partnership project (3GPP) completed the first version of the fifth-generation (5G) specifications in Release 15 in 2018 [1] and added additional functionalities to enhance performance and address new verticals in subsequent Releases 16 and 17 [2]. 3GPP Release 18 marks the start of 5G-Advanced, which represents a major evolution of 5G system and includes comprehensive work in the area of artificial intelligence (AI)/machine learning (ML) [3]. AI/ML can be used to manage complex network intelligently, address system optimization problems, and improve user experience in 5G system and beyond. In this article, we provide an overview of the AI/ML work in 3GPP Release 18, spanning across multiple 3GPP groups from services and system aspects (SA) to radio access network (RAN).

Before Release 18, 3GPP has carried out some initial work to embrace AI/ML techniques and data analytics in 5G system design. In Release 15, 3GPP introduced network data analytics function (NWDAF) in 5G core (5GC) network [4]. NWDAF can provide analytics to consumers such as 5GC network functions (NFs) and operations, administration, and maintenance (OAM). Since then, NWDAF functionality has been enhanced in subsequent releases. In particular, Release 17 specifies two NWDAF logical functions: Analytics logical function (AnLF) and model training logical function (MTLF). AnLF is responsible for inference, analytics information derivation, and analytics service exposure. MTLF is responsible for AI/ML model training and exposing new training services such as providing trained AI/ML models.

In the RAN domain, an initial study on AI-enabled RAN was completed in 3GPP Release 17 [5]. The study identified high-level principles and provided a functional framework for AI-enabled RAN intelligence. The study also investigated several use cases and solutions for AI/ML in RAN, including network energy saving, load balancing, and mobility optimization.

3GPP has also introduced management data analytics function (MDAF) as part of OAM [6]. MDAF, as a key enabler for network automation and intelligence, can process data related to network status and service events to provide analytics reports. The input data can come from different types of NFs (e.g., NWDAF) or entities (e.g., next generation node B (gNB)) in the network. MDAF can be deployed at different levels, including at domain level (e.g., RAN, core network) to provide domain-specific analytics and in a centralized manner to offer end-to-end or cross-domain analytics service.

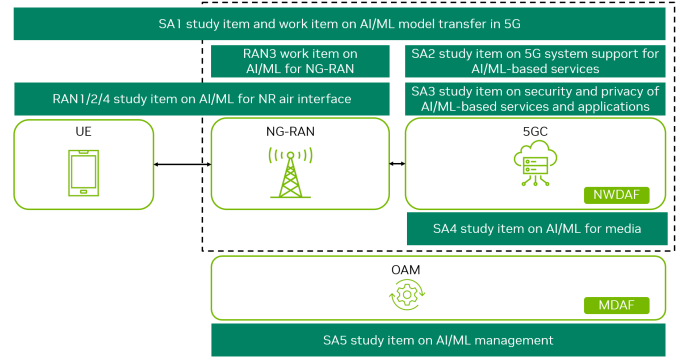

Figure 1: An overview of AI in 5G-Advanced in 3GPP Release 18 (indicative).

The 3GPP work on incorporating AI/ML techniques and data analytics into 5G system design conducted before Release 18 lays a solid foundation for further embracing AI/ML in 5G-Advanced evolution. Indeed, 3GPP Release 18 includes a diverse set of study and work items in the area of AI/ML, spanning across multiple 3GPP working groups, as illustrated in Fig. 1 and summarized below:

- SA working group 1 (SA1): Study item and work item on AI/ML model transfer in 5G.
- SA working group 2 (SA2): Study item on 5G system support for AI/ML-based services.
- SA working group 3 (SA3): Study item on security and privacy of AI/ML-based services and applications.
- SA working group 4 (SA4): Study item on AI/ML for media.
- SA working group 5 (SA5): Study item on AI/ML management.
- RAN working group 3 (RAN3): Work item on AI/ML for next-generation RAN (NG-RAN).
- RAN working groups 1, 2, and 4 (RAN1, RAN2, and RNA4): Study item on AI/ML for new radio (NR) air interface.

The AI/ML work for 5G-Advanced evolution in 3GPP Release 18 is anticipated to have a far-reaching impact on mobile technology evolution from 5G-Advanced to the sixth-generation (6G) mobile communications. It is critical to provide a holistic treatment of the AI/ML work in 3GPP Release 18 for the wireless communications and networking communities, which is the objective of this article. In the remaining of this article, we describe various 3GPP Release-18 study and work items on AI/ML as an organic whole, explain in detail the design aspects, and share various design rationales influencing standardization.

## II. APPLICATION LAYER AI/ML OPERATION IN 5G SYSTEM

### A. AI/ML Model Transfer

AI/ML models find application in mobile devices in 5G system, e.g., image recognition, speech recognition, and video processing. Due to diverse applications, changing environment, and limited storage at user equipment (UE), it is infeasible to load all AI/ML models in UE in advance. Thus, downloading AI/ML models as needed is a necessity. For some applications, UE may not have sufficient computation resource to perform inference, in which case the inference operation may need to be offloaded from UE to 5G cloud or edge. Besides, training data needs to be shared across different entities that collaboratively train a global AI/ML model in 5G system.

The trend of transferring AI/ML models and data brings in new traffic type that needs to be served by 5G system. 3GPP SA1 is responsible for identifying service and performance requirements for 3GPP systems. In Release 18, SA1 started a study to identify use cases and service and performance requirements for AI/ML model transfer over 5G system [7]. The study addressed three use cases which are illustrated in Fig. 2. The first use case is AI/ML operation splitting between AI/ML endpoints. The intention of the operation splitting is to keep privacy- or latency-sensitive parts of the operation in UE but offload the computation- or energy-intensive parts of the operation to network endpoints. The second use case is AI/ML model/data distribution and sharing over 5G system, whose purpose is to enable adaptive model downloading from a network endpoint to the devices when needed. The third use case is distributed/federated learning over 5G system, where UEs perform partial training based on local data and a central entity trains a global model by aggregating the local results from the UEs.

The study identified new service requirements and key performance indicators (KPIs) for AI/ML model training, inference, downloading, monitoring, prediction, and management over 5G system. After completing the study, 3GPP SA1 continued with a follow-up work item in Release 18 to specify service and performance requirements for 5G system to support AI/ML operations in the aforementioned three use cases. These requirements are captured in technical specification (TS) 22.261 [8].

From the service requirements' perspective, TS 22.261 specifies that for AI/ML model transfer, 5G system shall be able to support new service functionality related to, e.g., resource utilization monitoring, network performance monitoring, prediction and change notification, and quality-of-service (QoS) management. The applicability of the requirements is subject to operator policy, user consent, and regulatory requirements.

From the performance requirements' perspective, TS 22.261 specifies KPIs for AI/ML model transfer in 5G system, including end-to-end latency, experienced data rate, and

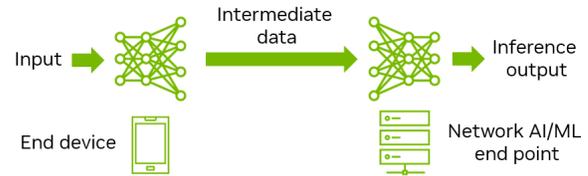

(a) AI/ML operation splitting between AI/ML endpoints

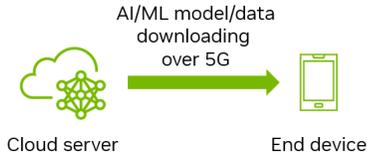

(b) AI/ML model/data distribution and sharing over 5G system

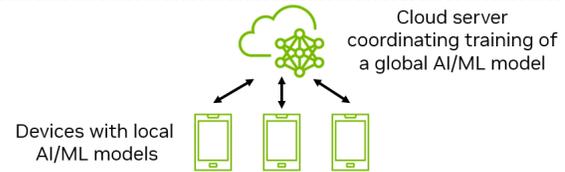

(c) Distributed/federated learning over 5G system

**Figure 2: Use cases of AI/ML operations in 5G system identified in 3GPP SA1 Release 18.**

communication service availability, among others. For example, the maximum allowed downlink end-to-end latency is 1 s and the experienced downlink data rate required is 1.1 Gbps for image recognition related AI/ML model distribution. As another example, in split AI/ML image recognition between UE and network server, the maximum allowed uplink end-to-end latency is 2 ms and the experienced uplink data rate required is 1.08 Gbps.

### B. 5G System Support

The 3GPP SA1 work on AI/ML model transfer in 5G system identified service and performance requirements, many of which are related to the support of application layer AI/ML operation in UE. Though 5G system includes NWDAF in 5GC to enable network automation, there are no optimized transport solutions specified for application layer AI/ML operation in UE. Therefore, it is necessary to evolve 5G system to support application layer AI/ML operation more intelligently. This consideration motivates the study item on 5G system support for AI/ML-based services [9], which is being conducted by 3GPP SA2, a working group responsible for developing the overall 3GPP system architecture.

From the perspective of architectural requirements, an application function (AF) controls the logic of the application layer AI/ML operation. AF is a control plane function which interacts with 5GC to provide services to support, for example, application influence on traffic routing. AF requests for 5G system assistance to support application layer AI/ML operation should be authorized by the 5GC.

3GPP SA2 is exploring several possible architectural and functional extensions to support application layer AI/ML operation. One key issue is about how to monitor network resource utilization related to the UE's performance such as data rate and latency. The network exposure function (NEF) in 5GC can support monitoring capability which allows configuration, detection, and reporting of monitoring events to authorized external party. New monitoring events may potentially be introduced to measure or predict network resource utilization for the support of application layer AI/ML operation. Besides, extending 5GC information exposure to UE or authorized third party may be beneficial in assisting the application layer AI/ML operation. The assistance information for exposure may include the UE or network conditions and performance predictions on, e.g., UE location, load, and QoS. Another key issue is about how to enhance provisioning capability which allows an external party to provision information to 5GC to facilitate the support of application layer AI/ML operation in 5G system. One example of the external parameter provisioning information is expected UE behaviors such as expected UE mobility and communication characteristics. It is also of interest to investigate other 5GC enhancements such as traffic routing mechanisms to better support AI/ML traffic transport.

5G system has comprehensive mechanisms for QoS definition, implementation, control, and monitoring to support various applications. Application layer AI/ML operation relies on the QoS provided by the underlying 5G system. It is necessary to study if the current 5G QoS model and policy framework need to be enhanced to account for AI/ML application traffic, expose monitoring and status information about network resource utilization to an authorized third party, and inform application layer AI/ML operation about predictions of changes in network conditions. Another objective of the 3GPP SA2 study is to investigate 5G system assistance to facilitate application layer federated learning operation over 5G system, including federated learning member selection and management, performance monitoring and exposure, and network resource allocation.

*C. Security and Privacy Aspects*

The 3GPP SA2 work on 5G system support for application layer AI/ML operation described in the previous section identifies that various AI/ML data needs to be transmitted between 5GC and AF. Some data (e.g., QoS analytics, geographical distribution information) may be user privacy sensitive and must not be exposed to an unauthorized AF. Some data (e.g., network load analytics) reflects network state and must be handled securely to prevent the data from being exploited by potential attackers. Therefore, it is critical to study security and privacy aspects of application layer AI/ML operation to avoid potential threat and risk to 5G system and users.

3GPP SA3, a working group responsible for security and privacy aspects in 3GPP systems, is performing a study on security and privacy of AI/ML-based services and applications in 5G [10]. The study aims to identify key issues, potential threats, requirements, and solutions. One identified key issue is about privacy and authorization for 5GC assistance information exposure to AF, including determination of privacy-sensitive assistance information and methods for 5GC to protect and authorize AF to access the information. Proper privacy protection and authorization mechanisms are needed to respect user privacy and ensure network security. One solution is to reuse existing mechanisms (e.g., open authorization (OAuth)) for authorization of 5GC assistance information exposure to AF.

Another solution under exploration in 3GPP SA3 is UE profile-based 5GC assistance information exposure authorization. A UE profile determines whether a specific AF can request or modify certain information of the associated UE. It may include UE identify, AF identity, expected service identifier, target 5GC assistance information, expiration time, and authorization and protection policies. The UE profile can be stored in unified data management (UDM)/unified data repository (UDR). An AF sends 5GC assistance information access request to NEF/NWDAF. Upon reception of the request, NEF/NWDAF directly determines the UE profile if already available in NEF/NWDAF; otherwise, NEF/NWDAF obtains the UE profile from UDM/UDR. Based on the UE profile, NEF/NWDAF checks whether the AF is authorized to access the 5GC assistance information. If so, NEF/NWDAF sends the 5GC assistance information with appropriate security protection mechanisms to the requesting AF.

## III. AI/ML FOR MEDIA

A vast amount of AI/ML applications are media related, such as image classification, speech recognition, and video quality enhancement. These applications require increasingly higher computational processing to handle the growing amount of AI/ML data and model complexity. Meanwhile, emerging media services such as virtual reality (VR), augmented reality (AR), and cloud gaming are taking off. Such mobile applications are processing intensive and heavily rely on AI/ML-assisted functions. For example, glass-based AR media services require context of the surrounding environment, which can be provided by using AI/ML-assisted object segmentation, recognition, and classification. AI/ML also finds application in media compression algorithms to make encoders more adaptive. Besides, the AI/ML models for the media applications require proper compression to facilitate efficient transmission. Indeed, the moving picture experts group (MPEG) has been working on deep neural network video coding to optimize media compression and on neural network representation to compress neural networks for media applications.

The aforementioned latest trends in the interplay between AI/ML and media applications call for an in-depth study to understand their impact on 5G system. Thus, 3GPP SA4, responsible for developing specifications of media codec and the system and delivery aspects of the media contents, is investigating AI/ML in 5G media services [11]. The main objective of the study is to identify relevant interoperability requirements and implementation constraints of AI/ML in 5G media services.

To start with, the 3GPP SA4 work aims to describe media-based AI/ML use cases and scenarios, based on the outcome of the 3GPP SA1 study on AI/ML model transfer that categorizes three key operations including AI/ML operation splitting, model/data distribution, and distributed/federated learning [7]. The first set of use cases are object recognition in image and video, which can be performed by UE inference only, network inference only, and split inferences. The second set of use cases are video quality enhancement in streaming. One example is end-to-end neural network-based video coding, where the sender processes a high-quality video using a neural network to generate an intermediate data stream and transmits it together with a lower resolution encoding of the video, and the receiver processes the received intermediate data stream and video stream to regenerate the high-quality video for rendering. One may also apply a neural network to postprocess a decoded video to enhance video quality. The third set of use cases are crowd-sourcing media capture, where each UE uses a shared neural network to process its captured video/audio, or the network performs inference on the media data from multiple UEs. The last set of use cases are natural language processing on speech, such as speech recognition and voice translation.

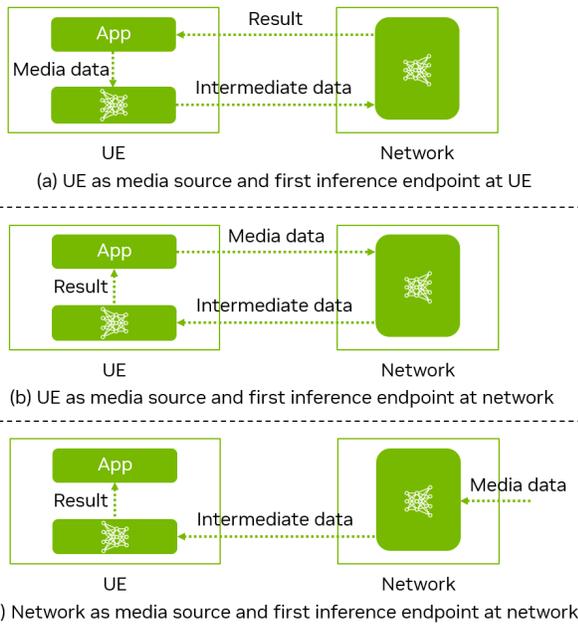

**Figure 3: Topologies of split AI/ML inference for media.**

A major objective of the 3GPP SA4 study on AI/ML for media is to describe media service architecture for AI/ML and relevant service flows. To this end, one key consideration is split AI/ML inference between network and UE. Split points can depend on a number of factors including UE capabilities (e.g., memory, compute, energy consumption, and inference latency), network conditions (e.g., capacity, load, and latency), model characteristics, and user/task specific requirements (e.g., delay and privacy). Figure 3 provides an illustration of different orders of operations and the corresponding media flows for split AI/ML inference between network and UE.

Other topics under exploration in the 3GPP SA4 study on AI/ML for media include data formats and protocols for various types of data components for AI/ML-based media services, traffic characteristics of the data components delivered over 5G, KPIs, and potential areas for normative work.

## IV. AI/ML MANAGEMENT IN 5G SYSTEM

AI/ML is being used across 5G system, including management and orchestration, core network, and RAN. To enable and facilitate AI/ML operation in 5G system, AI/ML models need to be created, trained, tested, deployed, and managed during its entire lifecycle. In Release 17, 3GPP conducted normative work on the management of AI/ML training [12] but the work did not cover other aspects such as AI/ML deployment and inference. Besides, AI/ML management capabilities may need to coordinate with AI/ML capabilities in core network and support AI/ML capabilities in RAN. 3GPP SA5, responsible for management, orchestration, and charging for 3GPP systems, is investigating AI/ML management to coordinate AI/ML functions across 5G system [13].

AI/ML operational workflow consists of three main phases: training phase (including model training and testing), deployment phase, and inference phase. AI/ML management for the training phase needs to support training data management, training management, testing management, and validation. AI/ML management for the deployment phase needs to support deployment control and monitoring, enabling the trained and tested AI/ML model to be deployed to the target inference function as well as monitoring the deployment process. AI/ML management for the inference phase needs to support activation and deactivation, inference function control, inference performance management, trustworthiness management, and inference orchestration.

3GPP SA5 has studied a comprehensive list of use cases for AI/ML management and their corresponding requirements and possible solutions. Use cases of management capabilities for the training phase include event data for training, entity validation, entity testing, entity retraining, joint training of entities, training data effectiveness reporting and analytics, context, entity capability discovery and mapping, update management, training performance evaluation, training configuration management, transfer learning, etc. Use cases of management capabilities for the inference phase include inference history tracking, inference orchestration, capability coordination, entity loading, inference emulation, inference performance evaluation, inference configuration management, update control, etc.

It is noted that trustworthiness is identified as a common management capability for both the training phase and the inference phase. The objective of trustworthiness management is to ensure that the AI/ML models are robust, explainable, and fair. The trustworthiness requirements may vary based on the risk level of the use case, requiring the trustworthiness mechanisms to be configured and monitored. As a first step, AI/ML trustworthiness indicators need to be defined. Depending on the use case, an AI/ML management service (MnS) consumer can select a proper set of trustworthiness indicators and request an AI/ML MnS producer to monitor and evaluate the selected indicators. Preprocessing of training/testing/inference data may be needed according to the desired trustworthiness measure of the corresponding AI/ML model. The AI/ML MnS should equip the consumer with the capability of providing trustworthiness requirements of data processing to the producer as well as enabling the producer to expose the supported trustworthiness-related data processing capabilities to the consumer. Similarly, the AI/ML MnS consumer should be able to query the AI/ML training producer, inference producer, and/or assessment producer about the supported trustworthiness capabilities and request the configuration, measurement, and reporting of a selected set of trustworthiness characteristics.

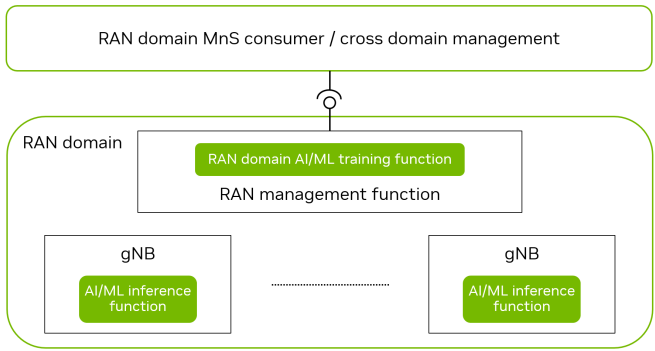

Figure 4: An example deployment scenario for RAN intelligence.

3GPP SA5 has also studied the deployment scenarios for AI/ML training function, testing function, inference function, and the corresponding management capabilities. The training/inference function can be located in cross-domain management system or domain management system (e.g., management data analytics in MDAF), core network (e.g., network data analytics in NWDAF), and RAN (e.g., RAN intelligence in gNB). The management capabilities provided by MnS producer are located in the corresponding management function (MnF). Different functions can be co-located or deployed in separate entities. Figure 4 illustrates a deployment scenario example for RAN intelligence, where AI/ML training function and inference function are located in RAN management entity and gNB, respectively, and the RAN management entity provides management capability for both the training function and the inference function.

## V. AI/ML FOR 5G RADIO ACCESS

RAN is arguably the most complex component of the cellular networks. In Release 18, 3GPP is engaging in normative work on AI-enabled RAN intelligence, as well as investigating the use of AI/ML to enhance the 5G NR air interface, to improve network performance and user experience.

### A. AI/ML-enabled NG-RAN

Following the completion of the Release-17 study [5], 3GPP RAN3, responsible for the overall RAN architecture and the specification of protocols for the related network interfaces, conducts normative work in Release 18 on AI-enabled RAN intelligence [14]. The objective of the normative work is to specify data collection enhancements and signaling support for AI/ML-based network energy saving, load balancing, and mobility optimization. The enhancements are introduced within the existing NG-RAN interfaces and architecture. Both non-split architecture (i.e., a monolithic gNB) and split architecture (i.e., gNB split into central unit (CU) and distributed unit (DU)) are in the scope of the work. AI/ML model training and model inference can be located in OAM and gNB (or gNB-CU for split architecture), respectively, or both functions can be located in gNB (or gNB-CU for split architecture).

Compared to the traditional measurement-based processes in the 3GPP specifications, AI-enabled RAN intelligence heavily relies on the use of predictions that can be exchanged among gNBs over Xn interface. Examples of prediction information include predicted cell-granularity UE trajectory and predicted resource status (e.g., predicted number of active UEs, predicted number of radio resource control (RRC) connections, and predicted radio resources). The cell-granularity UE trajectory prediction provides a list of cells where UE is expected to be located in future as well as the corresponding expected time of stay in each cell. Furthermore, feedback information that indicates how the RAN performance is affected by the AI/ML-based operations can be exchanged among gNBs. Examples of feedback information include UE performance (e.g., UE uplink/downlink throughput, packet delay, packet error rate) and energy efficiency metric.

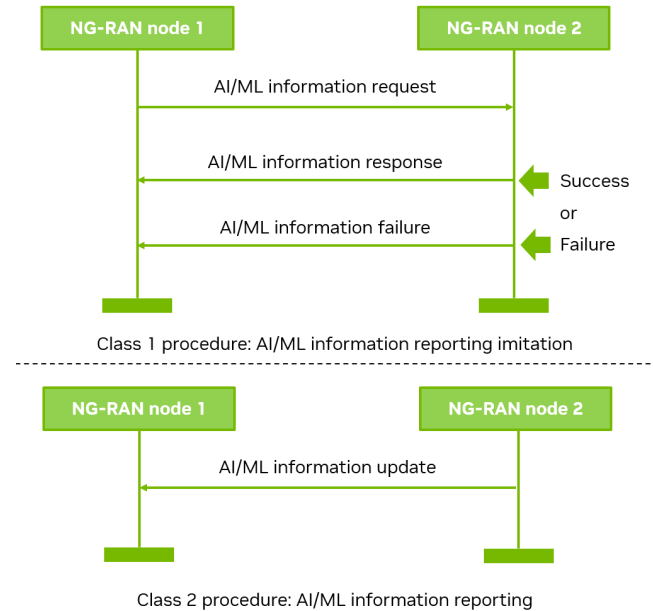

Figure 5: AI/ML information exchange procedure over Xn for NG-RAN.

To support the information exchange among gNBs for AI-enabled RAN intelligence, new signaling procedures are introduced for the Xn interface. There are two classes of elementary procedures for the Xn application protocol – Class 1: elementary procedures with response (success or failure) and Class 2: elementary procedures without response. The Class 1 procedure is used for AI/ML information reporting initiation, where an NG-RAN node requests the reporting of AI/ML related information from another NG-RAN node. The Class 2 procedure is used for AI/ML information reporting, where an NG-RAN node reports AI/ML related information after a successful AI/ML information reporting initiation procedure. These two classes of procedures are illustrated in Fig. 5.

### B. AI/ML for 5G Air Interface

AI-native air interface is anticipated to be key in 6G. To that end, it is crucial to act swiftly to investigate the use of AI/ML in 5G air interface already now to prepare for 6G standardization that will start in 3GPP around 2025. The 3GPP Release-18 study item on AI/ML for NR air interface marks a significant milestone in the evolution of cellular networks, as it is the first time such an approach has been taken. A main objective of this study item is to establish a general framework for enhancing the air interface using AI/ML, with various topics currently being explored. These include defining stages of AI/ML algorithms, collaboration levels between gNB and UE, required datasets for AI/ML model training, validation and testing, and life cycle management of AI/ML models.

There are three collaboration levels between gNB and UE under investigation in 3GPP:

- Level x: No collaboration.
- Level y: Signaling-based collaboration without model transfer.
- Level z: Signaling-based collaboration with model transfer.

Figure 6 provides an illustration of gNB-UE collaboration levels x, y, and z. At Level x, there is no collaboration between gNB and UE. The use of AI/ML techniques in this case is purely based on proprietary implementations and solutions. In particular, there is no dedicated AI/ML-specific enhancement (e.g., life cycle management related signaling) for AI/ML operations at Level x. At Level y, there is signaling-based collaboration without model transfer. Compared to Level x, the difference is that for Level y we may modify air interface and introduce new signaling to facilitate efficient AI/ML-based features, such as introducing new measurements and reporting. Level z is defined as signaling-based collaboration with model transfer, where model transfer refers to delivery of an AI/ML model over the air interface (either parameters of a model structure known at the receiving end or a new model with parameters) and the delivery may contain a full model or a partial model. It is noted that the boundary of Level y and Level z is defined based on whether model delivery is transparent to 3GPP signaling over the air interface or not. Specifically, model delivery in Level z is not transparent to 3GPP signaling, while Level y includes cases without model delivery and with model delivery transparent to 3GPP signaling over the air interface.

The study item is focused on three use cases serving as a pilot to deepen the understanding of the solution space through performance evaluation comparisons with pertinent non-AI/ML-based implementations.

*Channel state information (CSI) feedback*: The objective of this use case is to use AI/ML techniques to reduce CSI overhead, improve feedback accuracy, and enable prediction. One focused area is spatial-frequency domain CSI compression, which includes an AI/ML-based CSI encoder at UE and a corresponding AI/ML-based CSI decoder at gNB. Another focused area is time domain CSI prediction at UE, which uses an AI/ML model to infer future CSI based on historic CSI measurement results.

*Beam management*: The objective of this use case is to use AI/ML techniques to reduce beam management overhead and latency, as well as improving beam selection accuracy. One focused area is spatial-domain downlink beam prediction, which uses an AI/ML model to infer the best downlink beam in Set A of downlink beams based on the measurement results of Set B of downlink beams. Another focused area is time-domain downlink beam prediction, which uses an AI/ML model to infer the best downlink beam in Set A of downlink beams at a future time instant based on the historic measurement results of Set B of downlink beams.

*Positioning*: The objective of this use case is to use AI/ML techniques to improve positioning accuracy for different scenarios including, e.g., those with heavy non-line-of-sight (NLOS) conditions. One focused area is direct AI/ML positioning, which uses an AI/ML model to directly infer UE location (e.g., fingerprinting using channel observation as the input of the AI/ML model). Another focused area is AI/M assisted positioning, which uses an AI/ML model to infer an intermediate measurement statistic for positioning. Examples of the intermediate measurement statistics are line-of-sight (LOS)/NLOS probability, time-of-arrival, angle-of-arrival, and angle-of-departure.

To sum up, the comprehensive understanding of 3GPP's role in facilitating AI/ML incorporation in air interface obtained from the study will contribute to normative work not only in future releases of 5G-Advanced but also future generations of wireless systems developed by 3GPP, including 6G.

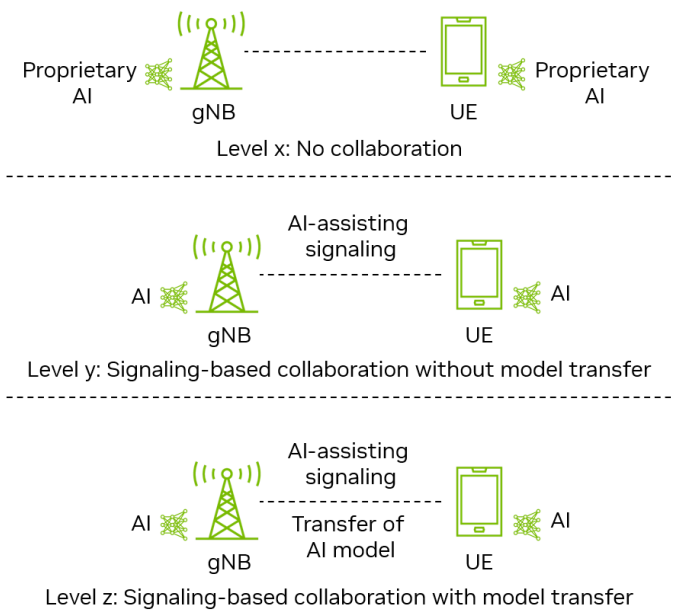

**Figure 6: Collaboration levels between gNB and UE in AI/ML for 5G air interface.**

## VI. CONCLUSION

Industries worldwide are being transformed by AI, and the telecom industry is no different. Standardization is crucial to facilitate widespread adoption of AI in telecom. 3GPP Release 18 – the first release of 5G-Advanced – includes a diverse set of study and work items dedicated to AI. This article has provided a holistic overview of the state of the art in the 3GPP work on AI in 5G-Advanced. We anticipate that 3GPP will continue fostering AI adoption from 5G-Advanced to 6G, triggering a paradigm shift in wireless communications and networking.